\documentclass[runningheads]{llncs}
\usepackage{graphicx,subcaption} 
\usepackage[linesnumbered,vlined,boxed,commentsnumbered]{algorithm2e} 
\usepackage[hidelinks]{hyperref} 
\usepackage[all]{hypcap} 
\usepackage[nameinlink,capitalise]{cleveref} 
\usepackage{xcolor} 
\usepackage{booktabs, makecell} 

\crefname{algocf}{alg.}{algs.}
\Crefname{algocf}{Algorithm}{algorithms} 

\title{OOPredictor: Predicting Object-Oriented Accesses using Static Analysis
\thanks{This research was conducted within the Centre for Advanced Studies-Atlantic
Faculty of Computer Science, University of New Brunswick.
The authors are grateful for the colleagues and facilities of CAS-Atlantic in supporting our research.
The authors would like to acknowledge the funding support of the Natural Sciences and Engineering Research Council of Canada (NSERC),
536287-18.
Furthermore,
we would also like to thank the New Brunswick Innovation Foundation for contributing to this project.}}
\titlerunning{OOPredictor}

\author{Hassan Arafat\inst{1}\orcidID{0000-0001-7148-0580} \and
  David Bremner\inst{1}\orcidID{0000-0003-0272-585X} \and
  Kenneth B. Kent\inst{1}\orcidID{0000-0003-2764-823X} \and
  Julian Wang\inst{2}\orcidID{0009-0009-7088-7827}
}

\authorrunning{H. Arafat et al.}

\institute{University of New Brunswick, Fredericton, Canada
  \and
  IBM Canada, Toronto, Canada
}

\begin{document}
\maketitle

\begin{abstract}
  Object-oriented Programming has become one of the most dominant design paradigms as the separation of concerns and adaptability of design reduce development and maintenance costs.
  However, the convenience is not without cost.
  The added indirection inherent in such designs causes excessive pointer chasing, negatively affecting locality, which in turn degrades the performance of cache structures.
  Furthermore, modern hardware prefetchers are mostly stride prefetchers that are ill-equipped to handle the unpredictability of access patterns generated by pointer chasing.
  Most software approaches that seek to address this problem resort to profiling the program as it runs, which comes with a significant run-time overhead or requires data from previous runs.
  In this paper, we propose the use of compile-time static analysis to predict the most common access patterns displayed by a program during run time.
  Since Java is one of the most popular object-oriented languages, we implement our prototype within the OpenJ9 JVM, inside the OMR optimizer infrastructure.
  The outputs of our proposed predictor are Markov chains that model the expected behavior of the program.
  The effectiveness of the proposed predictor is evaluated by comparing the model with the actual run-time behavior of the program measured using an instrumented interpreter.
  Our experiments show that the proposed predictor exhibits good accuracy and can be used to inform minimally intrusive load stall mitigation strategies,
  e.g.\ informing copying GCs on more locality-friendly copying orders.
\end{abstract}

\section{Introduction}
In modern computers, the main memory is much slower than the processing unit.
This gap in performance causes the memory to often be a performance bottleneck,
the effect is often referred to as the \emph{memory wall}.
In order to help bridge the gap,
computing systems utilize a hierarchy of progressively faster and smaller memories that
are used to retain copies of active memory locations,
those memories are often referred to as \emph{caches}.
Despite their small size compared to the main memory,
caches alleviate the impact of the memory wall significantly.
Their efficacy is driven by the fact that most programs' memory access patterns are not truly random;
they display some degree of temporal and spatial \emph{locality}.

To further improve the effectiveness of caches,
data can be loaded from the memory before they are needed, in what is known as \emph{prefetching}.
Most high-end modern processors have dedicated hardware units to perform prefetching.
Those \emph{hardware prefetchers} analyze the access pattern of the processor and prefetch data from the memory when they detect simple stride patterns.
While hardware prefetchers are effective at reducing cache misses from constant-stride accesses to arrays,
they struggle with more complex access patterns,
especially when the accessed addresses depend on values in memory.

A common memory access pattern that hardware prefetchers struggle with is called \emph{pointer chasing}.
Pointer chasing happens when the program has to retrieve the address of an object from memory by dereferencing a chain of pointers.
This negatively impacts cache performance as most production hardware prefetchers are not designed to handle pointer chasing.
The fact that the address to be accessed next is dependent on values in memory
means that the hardware prefetcher does not know the address to be loaded until it finishes loading from memory.
This results in a higher number of impactful load stalls and thus decreased performance~\cite{Indirect-Hot-Inoue}.
Prior knowledge of the program's access pattern could allow us to explore different mitigation strategies.

Pointer chasing patterns appear often in machine code generated from
\emph{Object-Oriented Programming} (OOP) languages.
OOP has become one of the most prevalent programming paradigms,
especially for enterprise solutions.
The division of code into manageable classes reduces development and maintenance costs.
This, however, comes at a performance cost as the abstraction requires added levels of indirection.
The added levels of indirection introduce extra pointer chasing.

In this paper we present a static analysis approach for predicting the pattern of object-oriented memory accesses in
programs written in Java. The work investigates the feasibility,
capabilities and limitations of such a system in a modern production runtime.
We also present a preliminary analysis on a possible use of the generated Markov chains to construct affinity graphs.

The rest of this paper is structured as follows:
we first discuss essential background concepts and elucidate related work,
then we describe our predictor,
followed by the methodology and experimental setup used to evaluate the predictor,
next we show data gathered from our experiment and our analyses
and finally we discuss our conclusions and future work.

\section{Background}
Programming is usually a challenging task,
especially when the programmer has to manage memory.
Modern programming languages, like Java,
offer a multitude of tools to help the programmer
and reduce development costs.

Those languages can also be run on complex runtimes that---in addition to garbage collection---include powerful tools to enhance performance,
collect information and provide interoperability.

\subsection{Managed Runtimes}
Managing memory is a tedious and error-prone process.
In order to simplify the job of a programmer,
managed languages unburden the programmer by automatically managing the memory;
they remove unused objects in what is known as \emph{garbage collection}.
While it does make the job of a programmer easier,
garbage collection introduces overhead that can degrade performance~\cite{GCImpact-Carpen-Amarie}.

Another one of the tools commonly available in modern runtimes is known as the \emph{Just In Time} (JIT) compiler.
It automatically and selectively compiles code on the run,
targeting the local hardware.
JIT compilers produce more optimal code---when compared to an interpreter---but they can also degrade performance if used haphazardly,
since they run during the execution of the program.
Virtual machines use complex policies to decide when and how to compile methods~\cite{JIT-Kulkarni},
in addition to a myriad of other cost reduction techniques,
e.g.\ Ahead-of-Time compilation~\cite{AOT-Krylov} and
disaggregated JIT servers~\cite{JITServer-Khrabrov}.

\subsection{Static Analysis}
To help improve performance,
optimizing compilers analyze the code to find opportunities for optimizations.
The analysis of the code itself before running is referred to as \emph{Static Analysis}.
On average, the more optimizations are enabled, the better-performing the resulting code and
the longer the compilation time.
The trade-off between compilation time and code performance is addressed by having multiple levels of optimization,
e.g.\ warm, hot, very-hot and scorching.
In a JIT environment,
methods recompile with progressively higher levels of optimization as they are invoked more frequently.

\subsection{Profiling}
Gathering data about the behavior of a program as it runs can be very helpful in guiding powerful optimizations.
This can be done in one of two approaches:
the data can be gathered by \emph{instrumenting} the program,
i.e., adding snippets of code into the program that collect statistics,
or by \emph{sampling} it with a different thread or process,
i.e., interrupting the execution of the program periodically and collecting information.
While the data gathered from these approaches can be useful,
the gathering of the data itself often incurs performance penalties.
Many approaches have been proposed to reduce this impact,
including running the profiler only for short bursts of time~\cite{Bursty-Hirzel}.

\subsection{OpenJ9}
\emph{Java} is a general purpose, object-oriented programming language.
Java source code is compiled into \emph{bytecode} that can be run on any machine with a Java Virtual Machine installed.
JVMs hide the complexities of the underlying machine and can run bytecode regardless of what type of machine produced it.

Eclipse OpenJ9~\cite{OpenJ9Website} is an open source JVM implementation
that utilizes Eclipse OMR's language-agnostic components~\cite{EclipseOMR}.
Those tools include garbage collectors,
and many compilation tools.

\section{Related Work}\label{sec:RelatedWork}
A recent study into the sources of hardware stalls in x86 architectures showed that object-oriented accesses are still a major source of load stalls in JVM workloads~\cite{StallBench-Zhuoran}.
There are many proposed optimizations that can use information about object access patterns of a program to help mitigate this issue and
those optimizations attempt to do that using a myriad of approaches.

Some approaches change the order of objects in memory,
while leaving their internal structure intact.
A main draw of such approaches is that they don't require any changes to the mutator.
Chen et al.~\cite{ProfileGC-Chen} propose a locality improving GC that can be triggered proactively to improve the cache performance of a program.
The locality improving GC uses profile information collected using read barriers and access sampling to identify hot objects.
During GC, those hot objects are copied first before processing the rest of the heap.
Another approach discussed by Serrano and Zhuang~\cite{DataContextGC-Serrano} uses the GC itself to build \emph{partial data context trees} (PDCT),
those trees are then combined with cache miss
information---collected by sampling hardware counters---to
model the object access pattern of a program.
They then use that information during copying in generational GC to better place objects in memory.
An innovative approach by Yang et al.~\cite{ZGCMutatorCopying-Yang} makes heavy use of access barriers in a region based GC to copy small objects in the order that the mutators access them.

Other approaches change the structure of the objects themselves.
Chilimbi, Davidson and Larus~\cite{ObjectRestructuring-Chilimbi} introduce a tool that can recommend better structure definitions for C programs.
The tool utilized both \emph{structure splitting}---splitting
the hot fields of structure so they can be collocated in a single cache
line---and \emph{field reordering}---changing
the order of fields in structures such that fields that have high temporal affinity are placed next to each other in memory,
using both dynamic information from a previous runs as well as static analysis of the program.
An approach proposed by Eimouri, Kent and Micic~\cite{JavaObjectSplitting-Eimouri} used \emph{object splitting} at allocation time in the JVM.\@
They use information collected using profiling to determine cold fields in objects,
the collected data is then used to allocate cold fields in separate regions of the \emph{balanced GC} policy in the OpenJ9 JVM. 

The previous approaches all used performance profiling during the run or required profiling data from previous runs.
The overhead of profiling as well as the impracticality of retaining profiling data were often a big factor in limiting the impact of those approaches.

Jeon, Shin and Han~\cite{StaticLayoutTransformations-Jeon} introduce the idea of using static analysis to predict accesses patterns and
then using that information to restructure the way objects in memory are allocated in C++.
Their predictions are rather simplistic and are represented as regular expressions,
which limits the amount of information the prediction provides.

Garbatov and Cachopo~\cite{AccessPatternAnalysis-Garbatov} present an analysis of different prediction models of database accesses of object-oriented applications,
they discuss \emph{Bayesian inference}, \emph{importance analysis} and \emph{Markov Chains}.
Their results show that \emph{Markov chains} is the model that shows least fluctuations.
However, the building and updating of their models is done using profiling data.

\section{Object-Oriented Access Predictor}
Our aim is to model the object-oriented behavior of the running program,
but without the overhead of fine-grain profiling or retaining previous profiles.
We thus propose a technique that uses \emph{static analysis} to build \emph{Markov chains} that model the object-oriented behavior of a program;
a novel static analysis-powered Object-Oriented access Predictor (OOPredictor).
OOPredictor can also be enhanced with branch bias information if they are available at compile time,
e.g., in a JIT environment.

The OOPredictor is implemented as an optimization within the OMR infrastructure used by the OpenJ9 JVM.
Considering OMR is language agnostic, it makes extending this work to other languages easier.
This is especially true for languages that have existing implementations that use OMR.\@
Another advantage of implementing our predictor as an optimization
is that it allows us to easily reuse the Control Flow Graph (CFG) that the compiler creates for other optimizations.
The JIT compiler works with the \emph{Intermediate Representation} (IR) of the instructions used by OMR,
known as \emph{TestaRossa Intermediate Language} (TRIL).

\subsection{Objective}
The predictor focuses mainly on making predictions about pointer-chasing patterns,
ones specifically stemming from the dereferencing object fields that are themselves object addresses.
The data gathered is class based,
informing us about the order that reference fields in an object of a specific class are likely to be accessed.
The fact that we ``reuse'' the CFG reduces the performance impact on compile time,
which allows for the predictor to be used in a JIT environment.
JIT compilation, however, does restrict the type of analysis that we can do to only intra-procedural analysis,
as more expensive analyses risk significantly increasing compilation time.

Those predictions, 
constructed without the use of profiling,
can be used to guide many optimizations that traditionally rely on profiling,
eliminating the costs associated with profiling.

\subsection{Output Format}
The output of the predictor can be described as a compressed Markov Chain,
where accesses that have 1.0 probability to be accessed in order are combined into one state.
We build nodes, each containing a list of fields that are expected to be accessed in-order.
Each node has a collection of outgoing edges to other nodes;
the weight of each edge represents the probability that the execution will transition to the destination node after finishing the current one.

\subsection{Building the Output}
To build the output Markov Chain we traverse the CFG.\@
For each \emph{basic block} (a portion of the code that has no branches and no branch targets)
we assume that accesses within are guaranteed to happen in-order---barring an interrupt,
and as such form the list of accesses in a single node.
This is assumed to hold since there cannot be any flow control within a \emph{basic block}
---by definition.
Each node can be thought of as a series of states that have a 100\% chance of transitioning from one to the next.

The fact we implement our approach in a JIT environment allows us access to some extra information,
e.g., statistics about branch biases as well some extra type information when polymorphism is present.
We can utilize the information to better inform our model
without incurring extra profiling costs as the information is collected irrespective of our optimization.
As a method is compiled to higher levels of optimization, 
multiple optimizations might cause changes to the resulting CFG and Markov chain.

In the case that such information is not available,
e.g., \emph{Ahead Of Time} (AOT) compilation,
we can set static probabilities to be given to forward and back branches.

Our main interest are IR instructions that are indirect accesses,
for each indirect access, we record the following:
\begin{enumerate}
  \item class: the class of the accessed field.
  \item field: the field within the object that is accessed.
  \item type: the class of the object that field references.
\end{enumerate}
The general outline of the algorithm is show in~\cref{alg:build_markov}.
The two loops in the beginning were implemented as one---with 
handling for yet to be constructed state,
but are separated into two distinct traversals in the pseudocode for clarity.

Transitions from a given block define which nodes are likely to be accessed next.
Since many blocks do not contain accesses that are of interest to us,
our resulting graph is full of empty nodes,
with only a few nodes that have any accesses at all.
Those empty nodes maintain only control flow information,
we want to retain that information but reduce the size of the output.
To do that, we \emph{bypass} all empty nodes except the nodes representing entry and exit blocks.

We traverse the graph and we \emph{bypass} nodes that are empty.
The operation \emph{bypass} removes the node and reconnects its parents to the children
while retaining the probabilities that each node is reached.
Self-loops, which can result from previous \emph{bypass} operations,
are removed and their probabilities distributed proportionally to outgoing edges.
This is possible because we only care about the control flow information in empty nodes.
Once the transitions are updated,
the bypassed node is removed~\cref{alg:bypass}.
\begin{algorithm}
  \capstart
  \AlgoDontDisplayBlockMarkers\SetAlgoNoEnd\SetAlgoNoLine%
  \SetKwData{CurrentMarkovState}{$\mathit{newstate}$}
  \SetKwData{TransitionWeight}{$\mathit{weight}$}
  \SetKwData{MarkovChain}{$\mathit{markovchain}$}
  \SetKwFunction{BreadthFirstTraverse}{breadthFirstTraverse}
  \SetKwFunction{IsOOAccess}{isOOAccess}
  \SetKwFunction{ToOOAccess}{toOOAccess}
  \SetKwFunction{AddAccess}{addAccess}
  \SetKwFunction{NewState}{newState}
  \SetKwFunction{FindState}{findState}
  \SetKwFunction{AddTransition}{addTransition}
  \SetKwFunction{IsBackwardEdge}{isBackwardEdge}
  \SetKwFunction{IsForwardEdge}{IsForwardEdge}
  \SetKwFunction{Bypass}{bypass}

  \SetKwInOut{Input}{input}
  \SetKwInOut{Output}{output}

  \Input{$\mathit{CFG}$: The CFG of the method getting compiled}
  \Output{\MarkovChain: A markov chain of OO Accesses}
  \BlankLine
  \tcp{Logic split into two loops for clarity}
  \ForEach{$\mathit{block}$ in \BreadthFirstTraverse{$\mathit{CFG}$}}{
    \CurrentMarkovState$\leftarrow$\MarkovChain.\NewState{$\mathit{block.number}$}\\
    \ForEach{$\mathit{instruction}$ in $\mathit{block.instructions}$}{
      \If{\IsOOAccess{$\mathit{instruction}$}}{
        \CurrentMarkovState.\AddAccess{\ToOOAccess{$\mathit{instruction}$}}
      }
    }
  }
  \ForEach{$\mathit{block}$ in \BreadthFirstTraverse{$\mathit{CFG}$}}{
    \CurrentMarkovState$\leftarrow$\FindState{$\mathit{block.number}$}\\
    \ForEach{$\mathit{successor}$ in $\mathit{block.successors}$}{
      \tcp{Use profiling values if available otherwise use defaults}
      \If{$\mathit{sucessor.frequecny}$ is available}{
        \TransitionWeight$\leftarrow$ $\mathit{successor.frequency}$
      }
      \Else{
        \tcp{static values depending on whether it is a forward edge or a backward edge}
        \TransitionWeight$\leftarrow$ static values}
    }
    \CurrentMarkovState.\AddTransition{\FindState{$\mathit{successor.target.number}$},\TransitionWeight}
  }
  \BlankLine
  \tcp{Call bypass on empty states}
  \ForEach{$\mathit{state}$ in \MarkovChain}{
    \If{$\mathit{state}$ is empty and $\mathit{state}$ is neither an initial nor final state}{
      \MarkovChain.\Bypass{$\mathit{state}$}
    }
  }

  \caption{Building Markov chain from CFG}\label{alg:build_markov}
\end{algorithm}

\begin{algorithm}
  \capstart
  \AlgoDontDisplayBlockMarkers\SetAlgoNoEnd\SetAlgoNoLine%
  \SetKwData{RedistributeWeight}{$\mathit{weight}$}
  \SetKwFunction{AddTransition}{addTransition}
  \SetKwFunction{DeleteTransition}{deleteTransition}

  \SetKwInOut{Input}{input}

  \Input{$\mathit{state}$: The Markov chain state being bypassed}
  \BlankLine
  \ForEach{$\mathit{edgein}$ in $\mathit{state.incoming}$}{
    \tcp{If it is a self-loop, redistribute probability equally}
    \If{$\mathit{edgein}$ in $\mathit{state.outgoing}$}{
      \RedistributeWeight${\leftarrow}$$\mathit{edgein.weight}/(\mathit{state.outgoing.size} - 1)$\\
      \DeleteTransition{$\mathit{edgein}$}\\
      \ForEach{$\mathit{edgeout}$ in $\mathit{state.outgoing}$}
      {
        $\mathit{edgeout.weight} = \mathit{edgeout.weight}+$\RedistributeWeight
      }
    }
    \Else{
      \ForEach{$\mathit{edgeout}$ in $\mathit{state.outgoing}$}
      {
        \tcp{add transition will just add to weight of existing transition if such a transition exists}
        $\mathit{edgein.source}$.\AddTransition{$\mathit{edgeout.target}$, $\mathit{edgein.weight} \times \mathit{edgeout.weight}$}
      }
    }
  }

  \caption{Bypassing an empty state}\label{alg:bypass}
\end{algorithm}

\section{Methodology}
Validating the output requires measuring the behavior of the JVM and comparing it with our models.
The instrumentation necessary for measuring is for evaluation purposes only and is not to be included in an actual production environment.

\subsection{Java interpreter instrumentation}
Our proposed predictor focuses solely on object-oriented accesses,
this precludes the use of a low-level profiler
as it would not be aware of object information in the JVM,
e.g., class names, field names, field types.
Consequently, we have decided to instrument the Java interpreter itself;
the interpreter can access object and class information easily,
and is relatively straight forward to instrument.
There are two relevant Java bytecodes that we will call \emph{object-oriented accesses}:
\begin{enumerate}
  \item getfield: retrieves the value of a field within an object.
  \item putfield: updates the value of a field within an object.
\end{enumerate}
For each object-oriented access we record the type and name of the field,
as well as the name of the class to which the field belongs.
We also save call sites to be able to identify the method model to compare against.

This approach has its challenges:
in order to record every access,
the programs had to be run in interpreter-only mode,
with option \emph{-Xint}.
For modern workloads,
interpreter-only execution time is---on
average---15 times longer than execution time with JIT enabled on the HotSpot JVM~\cite{Repositioning-Lambert}.
Similar results can be expected from the OpenJ9 JVM as neither machine is optimized for interpreted performance.
The slowdown is even greater when we take into account the instrumentation overhead.

Another problem is the size of the access logs
as they can get very large for bigger benchmarks.
Even with compression,
processing those files proved to be prohibitively expensive.
We consequently limited the size of the logs to $2\times10^9$ accesses.
This translates to multiple hours of runtime on average in our experiments,
giving the JVM more than ample time to reach a steady state.

It is important to note that while the instrumentation overhead would be prohibitively large in a production environment,
the instrumentation is only intended for the evaluation of the OOPredictor.
It is not part of the OOPredictor nor is it required by the OOPredictor to operate.

\subsection{Experimental Setup}
Every benchmark was run twice:
once with the JIT enabled to build the prediction model,
and once in interpreter-only mode to measure accesses.
Since we are measuring the behavior of the Virtual Machine (VM) itself,
not the underlying hardware,
there is negligible relevant variation between the two runs.
The measured accesses are compared to the model using a Python script,
the model is simulated like a non-deterministic state machine.
In order to maximize the amount of data,
we therefore add the predictor optimization to all optimization policies.

During the JIT run, the predictor is instructed to output the prediction model as a set of JSON files.
The files are then processed by the Python script and a list of symbol names for the relevant methods is saved.
The recorded accesses from the interpreter run are then processed and the call sites of the relevant methods are saved.
For each relevant method,
we iterate through its call sites and run the model against the recorded accesses.
For methods with more than 100 call sites, we select a random sample of 100 call sites.
Our reliance on the JIT compiler to make the models means that we only get models for the most executed methods.
Due to the limitation of the predictor to intra-procedural analysis,
we have to accommodate some ``gaps'' of accesses that are not predicted by our models.

For each method,
we record the number of accesses that were matched,
the number of accesses that were skipped and the number of calls where the predicted model finished execution in a final state,
i.e., one that represents an exit block in the CFG input.
We reiterate that the interpreter-only run and the offline processing here are only for this validation experiment,
and would not be part of a production use of the predictor.

We conclude the experiments with a brief case study of using the predictor to build affinity graphs for the chi-square benchmark from the Renaissance suite.
We built those affinity graphs by defining accesses that are within two blocks from each other as affine.
We compare those affinity graphs to those built more traditionally,
using a sliding a window of accesses recorded,
during an instrumented interpreter.
The list of relevant class names---those which the compiler generated an affinity graph for---is passed to interpreter.

As both set of graphs have the same nodes,
their adjacency matrices are of the same size and are then vectorized and compared directly.
It is important to note that since the affinity graph edges can only have positive values,
the resulting vectors are guaranteed to be in the positive orthant.
We use two metrics for measuring the similarity of the two generated graphs:
cosine similarity and Spearman's correlation coefficient.
Cosine similarity is a common way of quantifying the symmetry between vectors,
while we opt to use the Spearman's correlation coefficient as a way of measuring how the relative orders of edge weights compare.
We made that choice since the relative weight of edges is a major determining factor in many optimizations that use affinity graphs.

\subsection{Benchmarks}
We use the Renaissance benchmark suite~\cite{Renaissance-Prokopec} to evaluate our predictor.
The Renaissance benchmark suite covers a wide range of JVM workloads,
and has become one of the most ubiquitous modern suites.
We use all of the benchmark groups except the benchmarks in \emph{concurrent},
as they failed to run in interpreter-only mode because of timeouts.

We use \emph{SPECjbb2015} and \emph{SPECjbb2005} benchmarks to evaluate our predictor's performance on business workloads.
Although \emph{SPECjbb2005} has been retired,
its object access pattern is significantly different to that of its successor and our co-author informed us that it displays intriguing results in previous encounters.
SPECjbb2005 emulates server-side Java commonly used by businesses,
with clients and database replaced by driver threads and object binary trees, respectively~\cite{SPECjbb2005}.
SPECjbb2015 models the infrastructure of a worldwide supermarket company that handles point-of-sale requests,
online purchases and data mining~\cite{SPECjbb2015}.

\section{Results}
Since our approach is intra-procedural, 
we expect many of the accesses we can not match with our predictor to be a result of function calls.
So we allow for ``gaps'', 
during which accesses that are a result of another function are taking place and are thus not predicted by our Markov chains.
The accesses in those gaps are counted, but they will not cause the Markov chain to fail.
For each method we measure the percentage of calls that ended with the model in a final state---with
allowance for
gaps---we refer to that metric as the \emph{termination rate}.
To account for the impact of those gaps,
we also measure the percentage of object-oriented accesses that were correctly predicted by the model,
we refer to this metric as the \emph{OO match rate}.
The \emph{termination rate} is intended as a measure of how often an access sequence predicted by our model is observed during runtime.
On the other hand the \emph{OO match rate} is intended as a measure of how many of the object-oriented accesses that are measured were predicted by OOPredictor.

We also note the number of bytecodes within a method
as \emph{method size} and the number of OO accesses in the model as \emph{number of accesses}

We will use violin plots to visualize the data.
Violin plots show the arithmetic mean, median as well as the first and third quartiles.
They also show a Kernel Density Estimation (KDE) of the underlying data.
The KDE is built using a Gaussian kernel function and Scott's rule of thumb to select bandwidth~\cite{Density-Estimation-Scott}.

Since our data is non-parametric we use \emph{Spearman's rank correlation coefficient} to measure the following correlations:
\begin{itemize}
  \item \textbf{cot}: \emph{OO match rate} and \emph{termination rate}.
  \item \textbf{ctn}: \emph{Termination rate} and \emph{number of accesses}
  \item \textbf{con}: \emph{OO match rate} and \emph{number of accesses}
  \item \textbf{cts}: \emph{Termination rate} and \emph{method size}
  \item \textbf{cos}: \emph{OO match rate} and \emph{method size}
\end{itemize}

\subsection{SPECJBB}
\begin{figure}
  \centering
  \begin{subfigure}[t]{0.49\linewidth}
    \includegraphics[width=\linewidth]{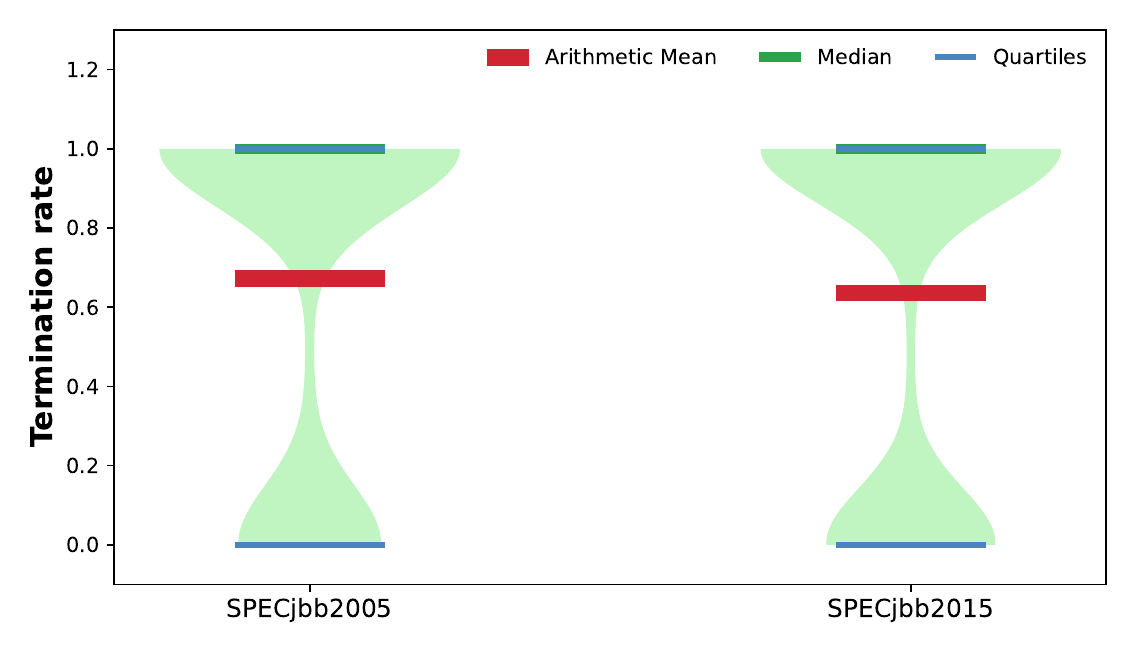}
    \caption{\emph{Termination rate} of SPECjbb workloads;
     the data is very top-heavy, this is clearly illustrated by the median and 3\textsuperscript{rd} quartile overlapping.}\label{fig:SPEC-graph-term}
  \end{subfigure}
  \hfill
  \begin{subfigure}[t]{0.49\linewidth}
    \includegraphics[width=\linewidth]{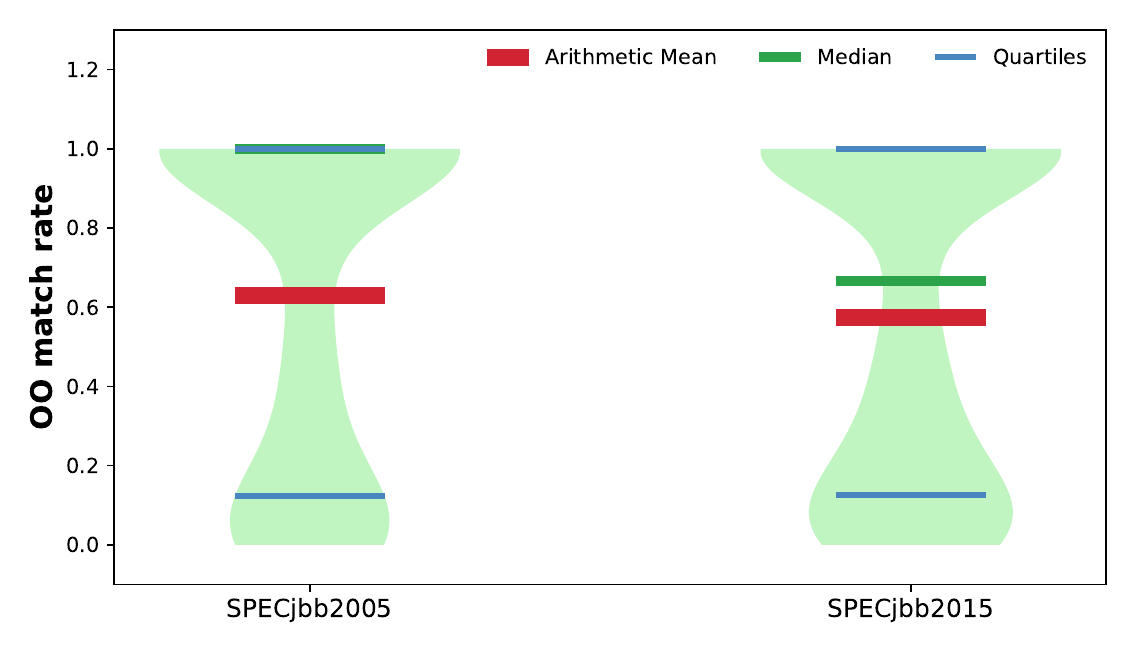}
    \caption{\emph{OO match rate} of SPECjbb workloads;
      the upper quartile is at the maximum as a quarter of the methods consistently achieve a value of almost 1.}\label{fig:SPEC-graph-oo}
  \end{subfigure}
  \caption{Violin plots for data collected from the SPECjbb workloads}
\end{figure}

The mean \emph{termination rate} for both \emph{SPECjbb2005} and \emph{SPECjbb\-2015} is around 0.7 as shown in \cref{fig:SPEC-graph-term}.
The KDE shape indicates the data is bimodal, with one peak around 1.0 and another around 0.0.
The median and 3\textsuperscript{rd} quartiles overlap at 1.0 indicating more than 50\% of the methods terminated on 100\% of the sampled sites.
The KDE of the \emph{OO match rate}, while showing the same bimodality,
shows larger spread of the values with some density in the interval 0.2--0.8 (\cref{fig:SPEC-graph-oo}).

\subsection{Renaissance}
Renaissance is a massive benchmark suite,
and as such we will address it in groups.
Benchmarks in the \emph{Concurrency} group could not be run,
as timeout exceptions kept occurring when running in interpreter-only mode.

\subsubsection{Apache Spark}
\begin{figure}
  \centering
  \begin{subfigure}[t]{0.49\linewidth}
    \includegraphics[width=\linewidth]{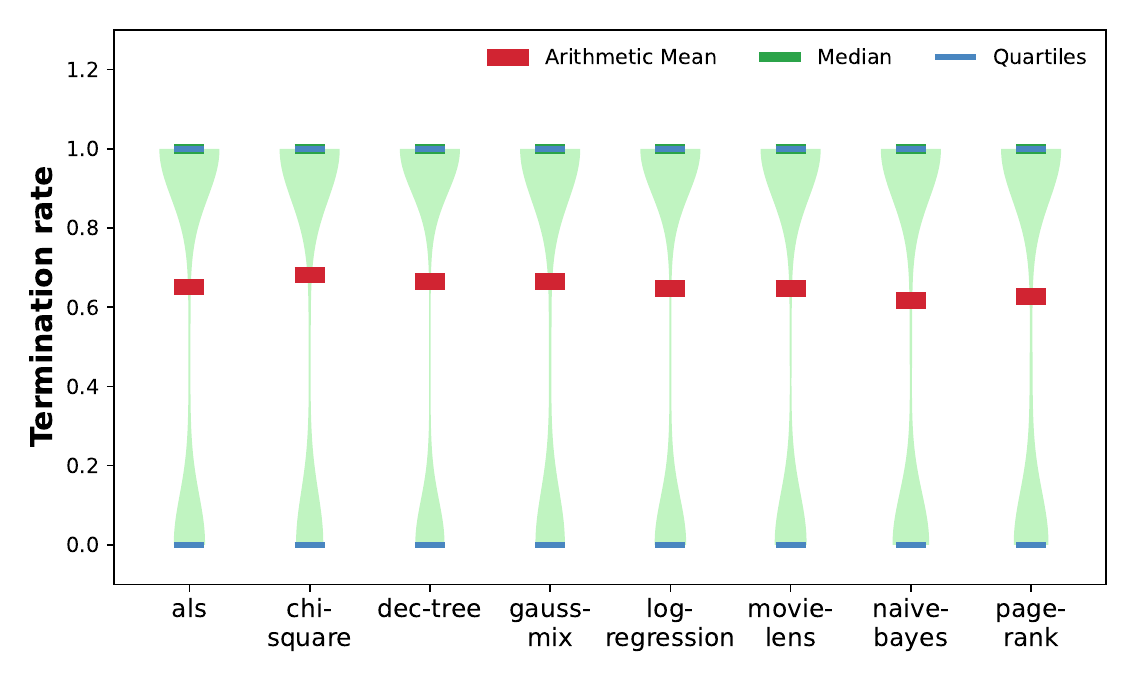}
    \caption{\emph{Termination rate} in the apache-spark group,
      the data is top heavy with the median and 3\textsuperscript{rd} quartile overlapping.}\label{fig:apache-spark-graph-term}
  \end{subfigure}
  \hfill
  \begin{subfigure}[t]{0.49\linewidth}
    \includegraphics[width=\linewidth]{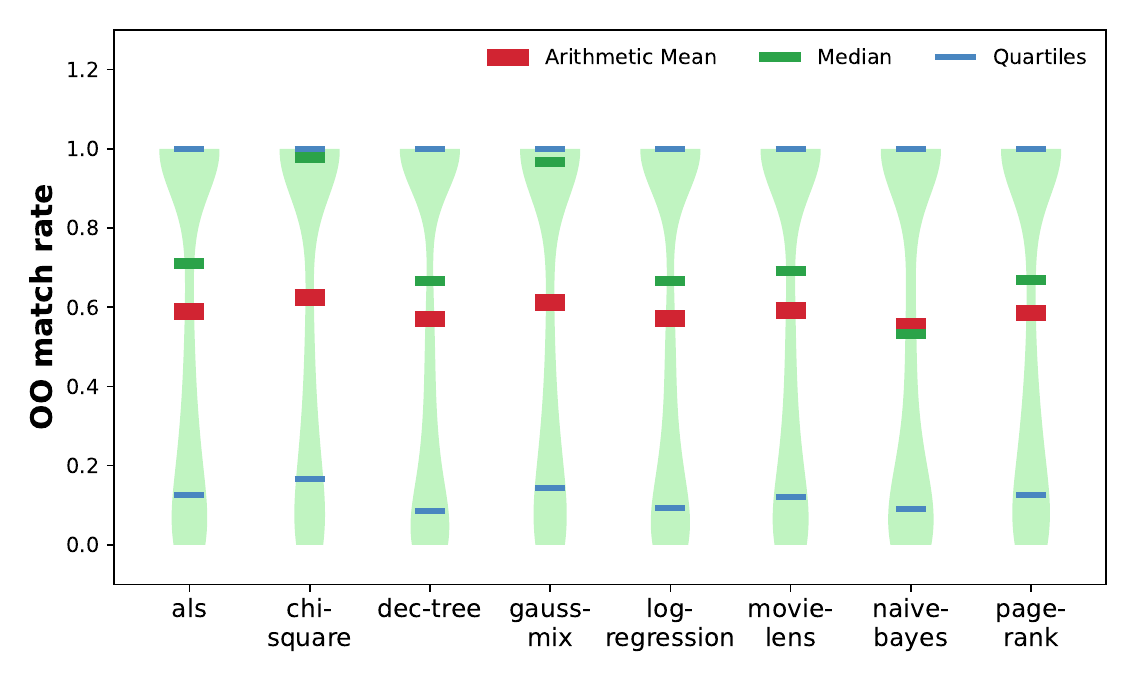}
    \caption{\emph{OO match rate} in the apache-spark group,
      the upper quartile is at the maximum as a quarter of the methods consistently achieve a value of almost 1.}\label{fig:apache-spark-graph-oo}
  \end{subfigure}
  \caption{Violin plots of data collected from the apache-spark group in the Renaissance suite}
\end{figure}

This group contains benchmarks that are compute-bound, data-parallel and focus on machine learning.
If we look at the \emph{Termination Rate}, shown in \cref{fig:apache-spark-graph-term},
we see that the data is extremely polarized.
The predictor reaches a final state in more than 80\% of the processed call sites for some methods,
while only doing so 20\% of time or less for others.
It is important to note that over 50\% of the methods in benchmarks of this group
have a \emph{termination rate} near 1.0,
causing the arithmetic mean and the 3\textsuperscript{rd} quartile to overlap at the 1.0.
While 25\% of the methods have a Termination rate near 0.0,
causing the 1\textsuperscript{st} quartile to be at the minimum.

The \emph{OO match rate} for the same group shows the same bimodal distribution,
but with more spread (\cref{fig:apache-spark-graph-oo}).
The medians for \emph{chi-square} and \emph{gauss-mix} are above 0.9,
meaning that 50\% of the methods in those benchmarks had a 90\% or more \emph{OO match rate}.

Method level correlations, shown in \cref{tab:apache-spark},
show only weak correlation between \emph{termination rate} and \emph{OO match rate}.
\emph{Termination rate} and \emph{OO match rate} are more strongly negatively-correlated
with the \emph{number of accesses} than raw \emph{method size}.

\subsubsection{Database}
We only use the \emph{db-shootout} benchmark,
as the \emph{neo4j-analytics} benchmark is not supported for JVM 11.
\emph{db-shootout} is a Java in-memory database benchmark that focuses on data structures and query-processing.
The data from \emph{db-shootout} shows more spread,
even for the \emph{termination rate} metric.
The arithmetic mean of both \emph{Termination rate} and \emph{OO match rate} were in the 0.5-0.6 range.

\subsubsection{Web}
This group contains two benchmarks:
\emph{finagle-chirper} and \emph{finagle-http}.
Both benchmarks simulate web servers with a focus on futures,
atomics and message-passing.
The \emph{termination rate} data displays strong bimodality with the arithmetic mean in the 0.7-0.8 range for the \emph{Termination rate} and in the 0.5-0.6 range for \emph{OO match rate}.

\subsubsection{Functional}
\begin{figure}
  \centering
  \begin{subfigure}[t]{0.45\linewidth}
    \includegraphics[width=\linewidth]{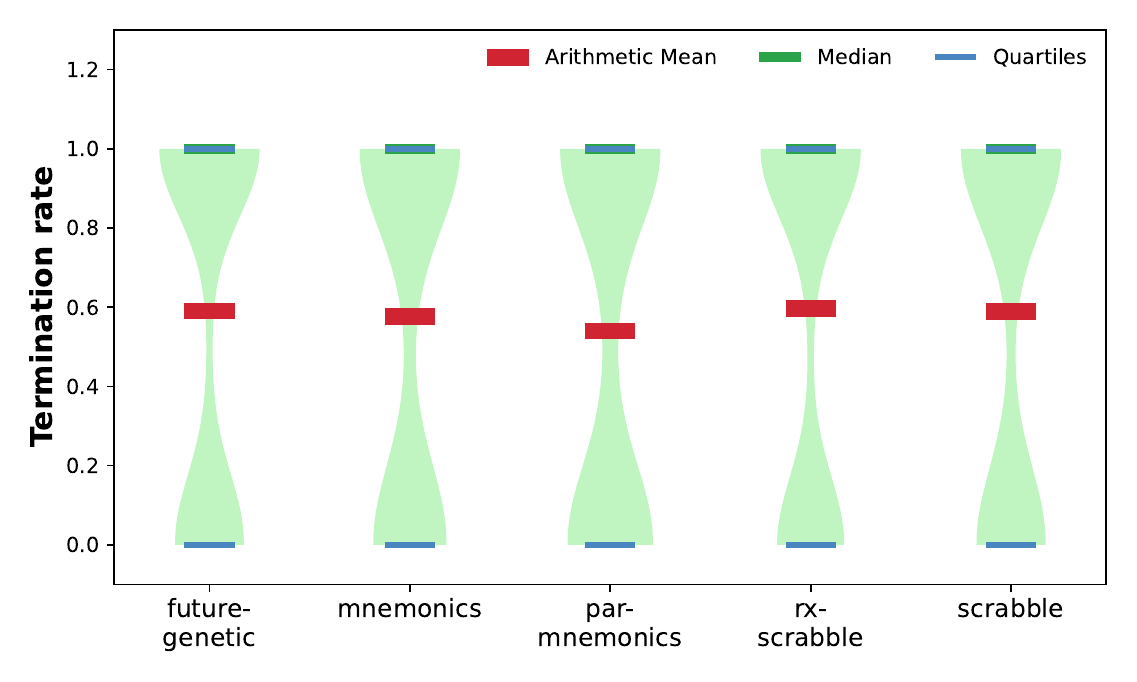}
    \caption{\emph{Termination rate} in the functional group,
      the data is top heavy with the median and 3\textsuperscript{rd} quartile overlapping.}\label{fig:functional-graph-term}
  \end{subfigure}
  \hfill
  \begin{subfigure}[t]{0.45\linewidth}
    \includegraphics[width=\linewidth]{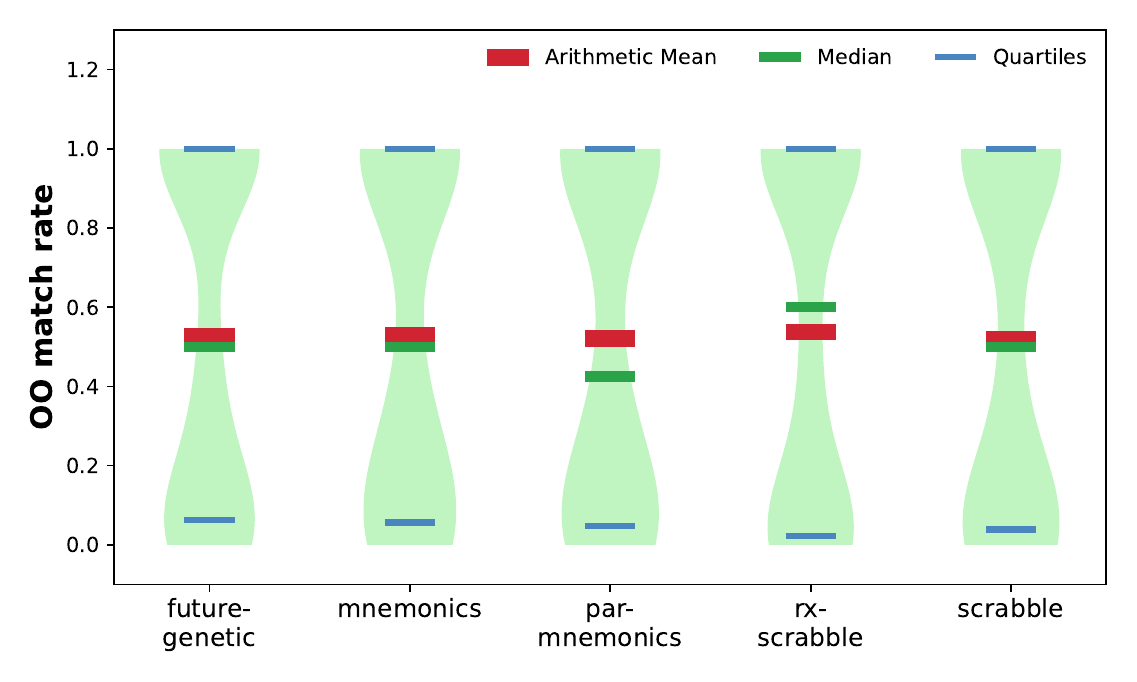}
    \caption{\emph{OO match rate} in the functional group,
      the upper quartile is at the maximum as a quarter of the methods consistently achieve a value of almost 1}\label{fig:functional-graph-oo}
  \end{subfigure}
  \caption{Violin plots of data collected from the functional group in the Renaissance suite}
\end{figure}

This group contains benchmarks that are task-parallel, memory-bound and have high contention.
Looking at the \emph{Termination Rate} graph,
we see that the arithmetic mean and 3\textsuperscript{rd} quartile overlap at 1.0,
this indicates that more than 50\% of the models from methods in these workloads always reached a final state (\cref{fig:functional-graph-term}).

The \emph{OO match rate} graph shows that the data is less polarized, with the KDEs getting wider near the center (\cref{fig:functional-graph-oo}).
The arithmetic means are also slightly lower than those in the \emph{apache-spark} group (\cref{fig:apache-spark-graph-oo}).
At the method level,
benchmarks in this group have a lower number of compiled methods and generally lower correlations across the board (\cref{tab:functional}).

\subsubsection{Scala}
This group features a collection of Scala benchmarks that use Software Transactional Memory (STM) or Scala collections.
The arithmetic means in this group are generally high, in the 0.6--0.8 interval.
This group has the highest arithmetic means out of any other group.
The patterns were very similar to benchmarks in the \emph{functional} group.

\begin{table}
  \parbox[t]{.49\linewidth}{
    \centering
    \caption{Table of the number of compiled methods and Spearman's rank correlation coefficients for workloads in the apache-spark group.}\label{tab:apache-spark}
    \resizebox{\linewidth}{!}{\begin{tabular}{c c c c c c c}
\toprule
\thead{}	& \textbf{methods}	& \textbf{cot}	& \textbf{ctn}	& \textbf{con}	& \textbf{cts}	& \textbf{cos}\\
\toprule
\textbf{als}	& 813	& 0.48	& -0.45	& -0.59	& -0.35	& -0.47\\
\midrule
\textbf{chi-square}	& 542	& 0.39	& -0.40	& -0.53	& -0.32	& -0.42\\
\midrule
\textbf{dec-tree}	& 1360	& 0.42	& -0.43	& -0.60	& -0.36	& -0.50\\
\midrule
\textbf{gauss-mix}	& 525	& 0.39	& -0.42	& -0.54	& -0.34	& -0.38\\
\midrule
\textbf{log-regression}	& 1126	& 0.42	& -0.43	& -0.59	& -0.34	& -0.48\\
\midrule
\textbf{movie-lens}	& 796	& 0.39	& -0.41	& -0.57	& -0.32	& -0.42\\
\midrule
\textbf{naive-bayes}	& 749	& 0.43	& -0.44	& -0.58	& -0.34	& -0.46\\
\midrule
\textbf{page-rank}	& 612	& 0.40	& -0.42	& -0.56	& -0.32	& -0.43\\
\bottomrule
\end{tabular}}
  }
  \hfill
  \parbox[t]{.49\linewidth}{
    \centering
    \caption{Table of the number of compiled methods and Spearman's rank correlation coefficients for workloads in the functional group.}\label{tab:functional}
    \resizebox{\linewidth}{!}{\begin{tabular}{c c c c c c c}
\toprule
\thead{}	& \textbf{methods}	& \textbf{cot}	& \textbf{ctn}	& \textbf{con}	& \textbf{cts}	& \textbf{cos}\\
\toprule
\textbf{future-genetic}	& 198	& 0.40	& -0.47	& -0.57	& -0.27	& -0.44\\
\midrule
\textbf{mnemonics}	& 89	& 0.39	& -0.43	& -0.48	& -0.34	& -0.34\\
\midrule
\textbf{par-mnemonics}	& 94	& 0.54	& -0.52	& -0.46	& -0.42	& -0.35\\
\midrule
\textbf{rx-scrabble}	& 173	& 0.38	& -0.41	& -0.34	& -0.25	& -0.24\\
\midrule
\textbf{scrabble}	& 105	& 0.30	& -0.45	& -0.44	& -0.33	& -0.23\\
\bottomrule
\end{tabular}}
  }
\end{table}

\subsubsection{Summary}
The bimodal nature of the collected data was consistent across benchmarks,
regardless of size of benchmark or nature of workload.
While the details of the distribution varied by workload,
the arithmetic mean values of both the \emph{termination rate}
and \emph{OO match rate}
remained consistently above 0.5.
The same can be said regarding the median,
with the sole exception of the \emph{par-mnemonics} benchmark in the \emph{Renaissance} suite.
Furthermore, 25\% of the models for every benchmark had a match rate of 100\% on both metrics.

Models that consistently have a high \emph{termination rate}
do not necessarily have a high \emph{OO match rate},
as the correlation coefficient between them is consistently less than 0.5.
Indicating that how often a predicted sequence appeared in-order had weak correlation to how many of the accesses that appeared were predicted.
Although the absolute value of the correlation coefficients remained under 0.65,
the size of the method as well as the number of object-oriented accesses in it were good indicators of lower predictor accuracy,
with the number of accesses showing a stronger correlation.
This was consistent across all benchmark groups, including those whose correlation data was omitted for brevity.

\subsection{Affinity Graphs}
Affinity graphs are used to guide many of the optimizations mentioned in~\cref{sec:RelatedWork},
but they are traditionally built using profiling.
We show a preliminary analysis of the accuracy of an affinity graph built from the chi-square benchmark of the Renaissance suite.
Two histograms show the \emph{cosine similarity} and \emph{Spearman's correlation coefficient} between the class affinity graphs
built using models produced from the OOPredictor and those produced traditionally using an instrumented interpreter.

\begin{figure}
  \centering
  \begin{subfigure}[t]{0.45\linewidth}
    \includegraphics[width=\linewidth]{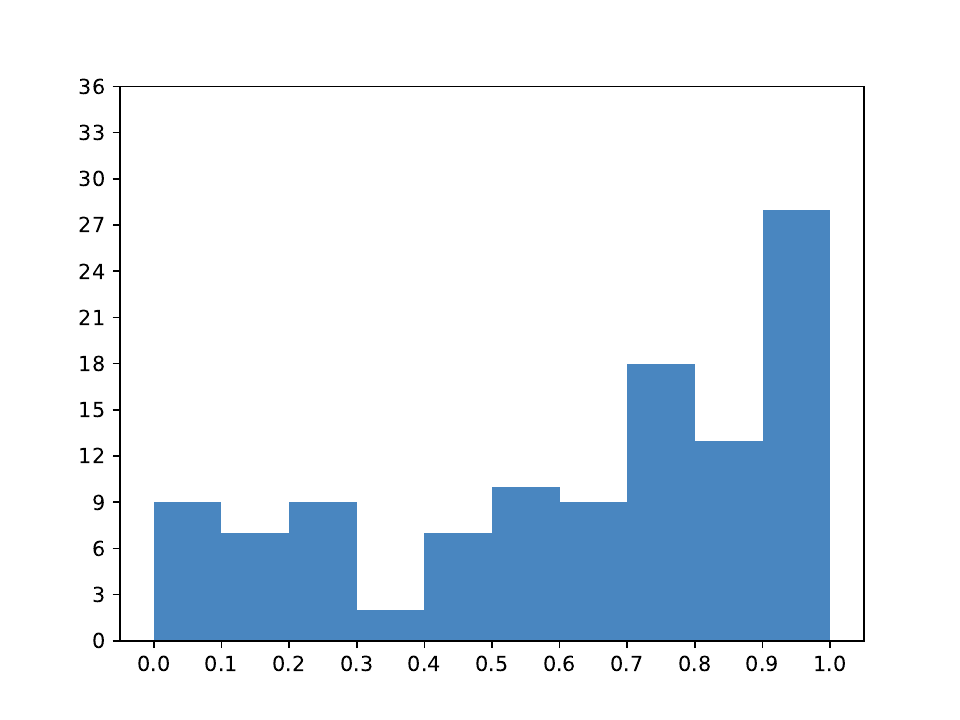}
    \caption{Histogram of \emph{cosine similarity} between the affinity graphs generated using \emph{OOPredictor} and the ones built via profiling.}\label{fig:cos_aff_graph}
  \end{subfigure}
  \hfill
  \begin{subfigure}[t]{0.45\linewidth}
    \includegraphics[width=\linewidth]{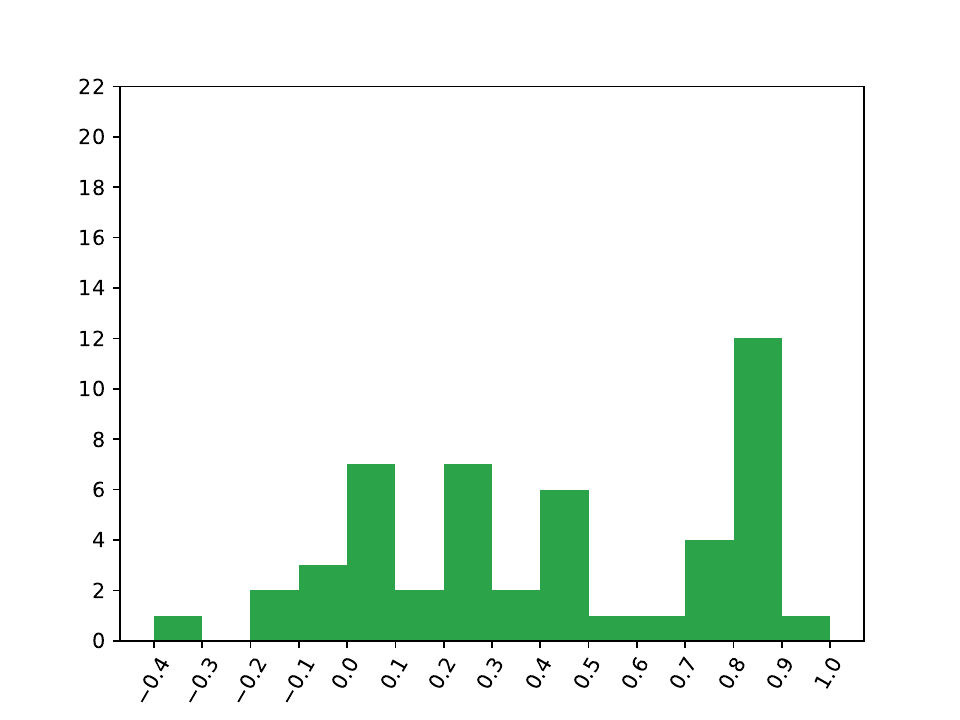}
    \caption{Histogram of \emph{Spearman's coefficient} between the affinity graphs generated using \emph{OOPredictor} and the ones built via profiling.}\label{fig:spear_aff_graph}
  \end{subfigure}
  \caption{Similarity metrics between affinity graphs from the chi-square benchmark}
\end{figure}

For Spearman's correlation,
we only consider classes where the Spearman's correlation test that meet our significance level requirement,
i.e., $p-value \leq 0.05$.
The distribution of the cosine results is skewed sharply to the right.
The highest number of classes is in the 0.9--1.0 range,
and the second highest in the 0.7--0.8 range,
with the rest of the domain having values distributed roughly evenly (\cref{fig:cos_aff_graph}).
Furthermore, the Spearman's coefficient results show that the relative order of edges is strongly correlated for over a third of the classes,
with a similar number of classes showing weak correlation (\cref{fig:spear_aff_graph}).
 
\section{Conclusion}
The proposed predictor is consistently accurate for certain methods
while frequently failing to make accurate predictions for others.
The size of the resulting model is a better indicator of predictor accuracy than plain method size,
which would encourage us,
in situations where we would like to limit our predictor to ``easily predictable'' methods,
to at least preliminarily determine the number of OO accesses in a method before deciding if we should use the predictor.

It is important to note that since the predictor models fail to predict the object-oriented access patterns of some functions,
it is best to use this information to guide non-intrusive and low risk optimization strategies to improve the layout of objects in memory and thus the spatial locality of the programs.

We believe the results show that the proposed predictor can provide useful information about the runtime object-oriented access pattern of a program at the compilation time of its methods.
Without having to incur the cost of extra runtime profiling.
The models can be used to extract class affinity information about the program at compile time which can then be used by optimization strategies either in the compiler or the GC to improve performance.
Our preliminary affinity graph analyses show the approach has promise.
As for future work, further study into the feasibility of expanding this approach to inter-procedural analysis and how it affects both cost and accuracy of the predictor remains to be done.
The affinity graph analysis needs to be expanded to other benchmarks,
as well as improved metrics and sensitivity analyses.
Affinity graphs built at compile time can be used to inform field layout optimizations at object allocation time,
as well as object layout optimizations.

\bibliographystyle{splncs04}
\bibliography{OOPredictor}
\end{document}